\documentclass[12pt,twoside]{article}
\usepackage{fleqn,espcrc1}

\usepackage{graphicx}
\usepackage[figuresright]{rotating}

\newcommand{\AmS}{{\protect\the\textfont2
  A\kern-.1667em\lower.5ex\hbox{M}\kern-.125emS}}

\title{NEW DATA FROM SND DETECTOR IN NOVOSIBIRSK} 
\author{
M.N.~Achasov, V.M.~Aulchenko,  K.I.~Beloborodov, A.V.~Berdyugin,\\
A.V.~Bozhenok, A.D.~Bukin, D.A.~Bukin, S.V.~Burdin, T.V.~Dimova,\\
S.I.~Dolinsky, A.A.~Drozdetsky, V.P.~Druzhinin, M.S.~Dubrovin, 
I.A.~Gaponenko,\\
V.B.~Golubev, V.N.~Ivanchenko, I.A.~Koop, A.A.~Korol, M.S.~Korostelev,\\
S.V.~Koshuba, G.A.~Kukartsev, E.V.~Pakhtusova, V.M.~Popov, A.A.~Salnikov,\\
S.I.~Serednyakov, V.V.~Shary, Yu.M.~Shatunov, V.A.~Sidorov, Z.K.~Silagadze,\\
A.N.~Skrinsky, Yu.V.~Usov,Yu.S.~Velikzhanin, A.V.~Vasiljev, A.S.~Zakharov\\
{\em Budker Institute of Nuclear Physics, Novosibirsk 630090, Russia,}\\
{\em Novosibirsk State University, Novosibirsk 630090, Russia}\\
{ ~}\\
{presented by V.P.~Druzhinin}
}
\begin{document}
% typeset front matter
\maketitle
\begin{abstract}
  The current status of experiments with SND detector at VEPP-2M
$e^+e^-$ collider in the energy range $2E_0=400-1400$ MeV and
recent results of data analysis for $\phi$, $\omega$ and $\rho$ decays
and $e^+e^-$ annihilation into hadrons are presented.
\end{abstract}
\section{Introduction}
VEPP-2M is the $e^+e^-$-collider\cite{VEPP2M}, operating
since 1974 in the energy range 2E=0.4-1.4 GeV
($\rho,\omega,\phi$-mesons region). Its maximum luminosity is about
$5\cdot10^{30}\,cm^{-2}s^{-1}$ at E=510 MeV. Two detectors
SND and CMD2 carry out experiments at VEPP-2M now.

  SND was described in detail in \cite{SND}.
Its main part is the three layer spherical electromagnetic calorimeter 
consisting of 1620 NaI(Tl) crystals with total mass of 3.6 tones. 
The solid angle coverage of the calorimeter is 90\% of $4\pi$ steradian. 
The energy resolution
for photons can be approximated as $\sigma_E(E)/E=4.2\%/E(GeV)^{1/4}$,
angular resolution is about $1.5^{\circ}$. The angles of charged
particles are measured by two cylindrical drift chambers covering 95\%
of $4\pi$ solid angle.
  
  In this report we present results of the four SND experiments:\\
\textbullet {\bf PHI96}:
7 scans of energy region 985--1040~MeV were performed with total integrated
luminosity $4.3~pb^{-1}$. Number of produced $\phi$-mesons is about 
$8\cdot10^6$.\\
\textbullet {\bf PHI98}:
$8~pb^{-1}$ of integrated luminosity were collected in two scans of the 
energy region 985--1060~MeV. Number of produced $\phi$-mesons is about 
$12\cdot10^6$.\\
\textbullet {\bf MHAD97}: the data was collected
in the energy region 1040-1380~MeV with the
integrated luminosity $6.3~pb^{-1}$.\\
\textbullet {\bf OME98}: the data was collected
in the energy region 360-970~MeV with the
integrated luminosity $3.6~pb^{-1}$. $1.2\cdot10^6$ $\omega$-mesons
and $1.9\cdot10^6$ $\rho$-mesons were produced. 
%\end{itemize}
\section{$\phi$-meson decays}
\subsection{Magnetic dipole decays $V\to P \gamma$}
The radiative decay $\phi\to\eta\gamma$ was studied
in three main $\eta$ decay modes. The obtained results are
the following:\newline
$$Br(\phi\to\eta\gamma)=\left\{ 
\begin{array}{lll}
(1.296\pm 0.024\pm 0.057)\%, & \eta\to 3\pi^0          & \mbox{\cite{3pi0}} \\
(1.338\pm 0.012\pm 0.052)\%, & \eta\to 2\gamma         & \mbox{\cite{2gam}} \\
(1.259\pm 0.030\pm 0.059)\%, & \eta\to\pi^+\pi^-\pi^0  & \mbox{\cite{3pi}}  \\
\end{array} \right.$$
The measured branching ratios are consistent inside systematic errors.
The accuracy of our average
\boldmath$Br(\phi\to\eta\gamma)=(1.304\pm 0.049)\%$ \unboldmath is better 
then world average one\cite{PDG}.

 The $\phi\to\pi^0\gamma$ decay was studied in 3-gamma final state.
The result \boldmath$B(\phi \to \pi^0\gamma )=
(1.226\pm 0.036_{-0.089}^{+0.096})\cdot 10^{-3}$\unboldmath\cite{2gam} is
most precise measurement of the $\phi\to\pi^0\gamma$ decay. The main problem of
this measurement is the large systematic error coming from
uncertainty in the relative $\phi$-$\omega$ phase. The improvement 
can be achieved with the precise measurement of non-resonant cross section
in the energy region between $\omega$ and $\phi$.
\subsection{$\phi\to\pi^0\pi^0\gamma, \eta\pi^0\gamma$ decays}
First observation of the $\phi\to\pi^0\pi^0\gamma, \eta\pi^0\gamma$ decays
was reported in 1997 by SND.\cite{hadron97}. Results of completed analysis
of PHI96 experiment were published in \cite{pi0pi0gam,etapi0gam}. These
results were confirmed by CMD-2 \cite{CMD1}. The $\phi\to\pi^0\pi^0\gamma,
\eta\pi^0\gamma$ decays are interesting for clarification of the nature
of $f_0(980)$ and $a_0(980)$ mesons. The theoretical estimations of
$\phi\to f_0\gamma, a_0\gamma$ branching ratios vary from about $10^{-5}$
in two-quark and $K\bar{K}$ molecular models up to $10^{-4}$ in four-quark
model\cite{4quark}.

 We studied the $\phi\to\pi^0\pi^0\gamma, \eta\pi^0\gamma$ decays in five
photons final state. The energy-momentum balance and existence of
$\pi^0\pi^0$ or $\eta\pi^0$ intermediate state were required for each
events. Additional cuts were imposed to suppress background from
$e^+e^-\to\omega\pi^0$ and $\phi\to\eta\gamma\to\ 3\pi^0\gamma$ processes.
The detection efficiencies for both decays under study strongly depended
on recoil photon energy. The efficiency averaged over photon spectrum
was about $15\%$ for $\pi^0\pi^0\gamma$ and $3.5\%$ for
$\eta\pi^0\gamma$ events.
\begin{figure}[tb]
\begin{minipage}[t]{0.46\textwidth}
\includegraphics[width=0.9\textwidth]{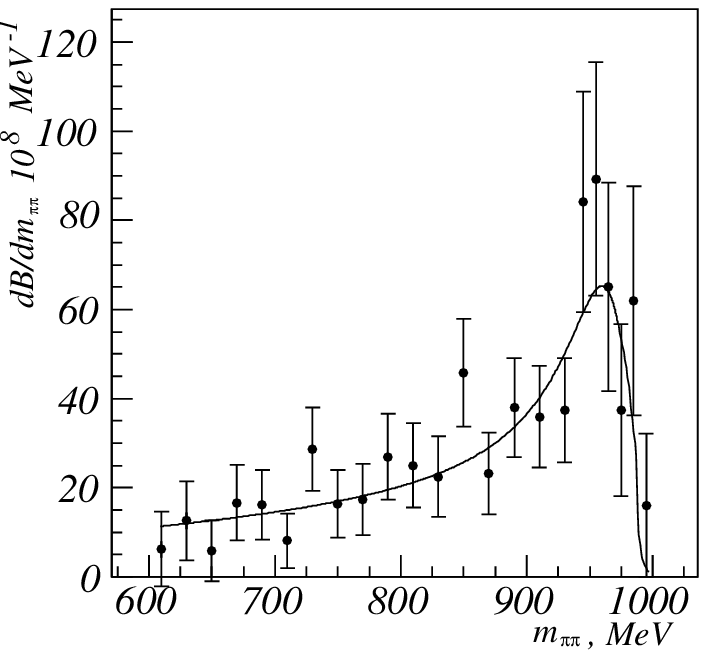}
\caption{The $\pi^0\pi^0$ invariant mass spectrum for
$\phi\to\pi^0\pi^0\gamma$ decay.}
\label{fig1}
\end{minipage}
\hspace{\fill}
\begin{minipage}[t]{0.46\textwidth}
\includegraphics[width=0.98\textwidth]{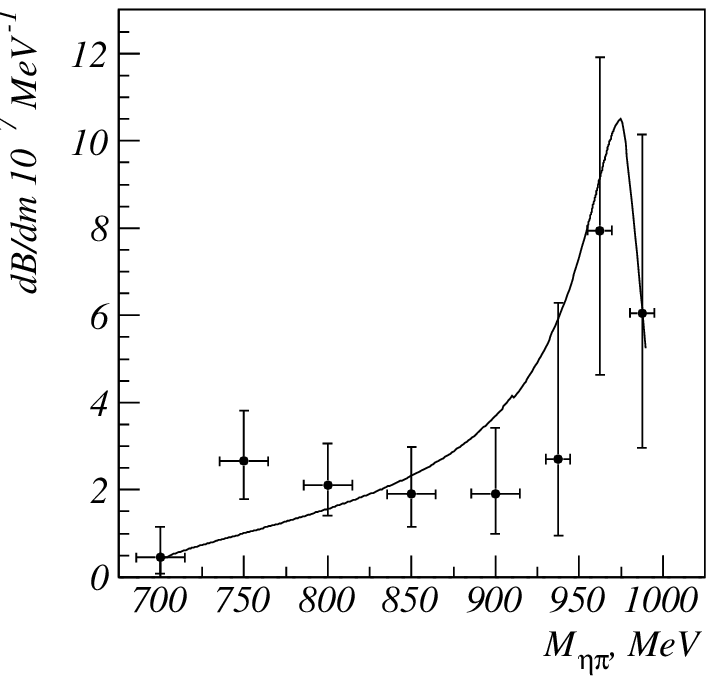}
\caption{The $\eta\pi^0$ invariant mass spectrum for
$\phi\to\eta\pi^0\gamma$ decay.}
\label{fig2}
\end{minipage}
\end{figure}

The $\pi^0\pi^0$ and $\eta\pi^0$ mass spectra are presented in Fig.\ref{fig1},
\ref{fig2}. For both spectra significant part of events is placed at large
masses supporting assumption about $f_0(980)$ and $a_0(980)$ intermediate
states. The obtained branching ratios for $\phi\to\pi^0\pi^0\gamma$ decay
\cite{pi0pi0gam} are
\begin{eqnarray*}
B(\phi\to\pi^0\pi^0\gamma)=(1.14\pm0.10\pm0.12)\cdot 10^{-4}, &\\
B(\phi\to\pi^0\pi^0\gamma)=(0.50\pm0.06\pm0.06)\cdot 10^{-4}, &
\mbox{for  } m_{\pi\pi}>900\,MeV
\end{eqnarray*}
This result is based on analysis of only PHI96
experiment. For $\phi\to\eta\pi^0\gamma$ decay we present the preliminary
result of analysis of full $\phi$ meson data sample:
\begin{eqnarray*}
B(\phi\to\eta\pi^0\gamma)=(0.87\pm0.14\pm0.07)\cdot 10^{-4}, &\\
B(\phi\to\eta\pi^0\gamma)=(0.36\pm0.11\pm0.03)\cdot 10^{-4}, &
\mbox{for  }m_{\eta\pi}>950\,MeV.
\end{eqnarray*}
The new branching ratio value is in
good agreement with our previous result $(0.83\pm0.23)\cdot
10^{-4}$\cite{etapi0gam}, but its statistical error is two time smaller.

Our conclusions on $\phi\to\pi^0\pi^0\gamma$ and $\phi\to\eta\pi^0\gamma$
decays are that greater part of events for both decays originate
from $f_0\gamma$ or $a_0\gamma$ intermediate states. The relatively large 
values of measured branching ratios contradict conventional two-quark 
structure of $f_0(980)$ and $a_0(980)$ mesons, but support their exotic 
four-quark structure.
\subsection{$\phi\to\omega\pi$ and $\phi\to\pi^+\pi^-$ decays}
These double suppressed by QZI rule and G-parity decays can be seen
as interference patterns in the cross sections of $e^+e^-\to\omega\pi$ and
$e^+e^-\to\pi^+\pi^-$ processes. The Born cross section for such processes
is described the following formula:
$$\sigma(E) =\sigma_0(E)\cdot
\biggl |1-Z \frac{m_\phi\Gamma_\phi}{D_\phi(E)}\biggr |^2,$$
where $\sigma_0(E)$ is nonresonant cross section, $Z$ is complex
interference amplitude, $D_\phi(E)$ is $\phi$ meson propogator.
One can extract from experimental data both real and imaginary parts 
of the decay amplitude.
\begin{figure}[tb]
\begin{minipage}[t]{0.46\textwidth}
\includegraphics[width=0.9\textwidth]{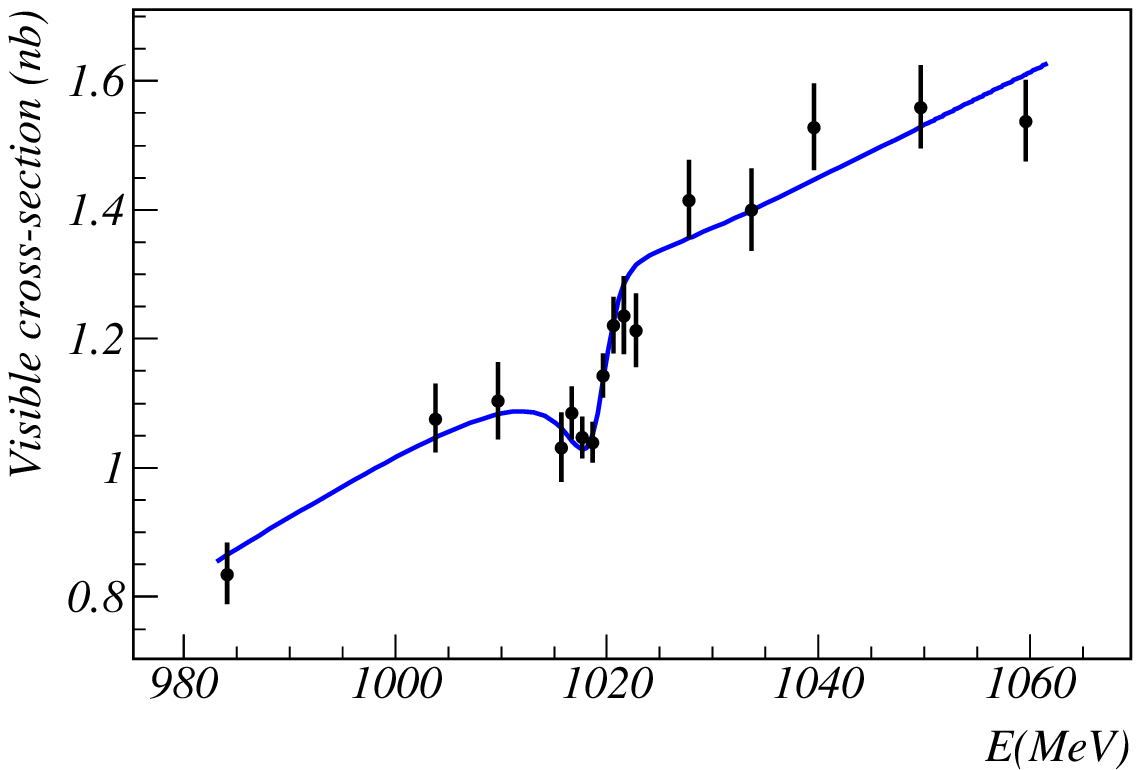}
\caption{The visible cross section of
$e^+e^-\to\omega\pi^0\to\pi^+\pi^-\pi^0\pi^0$ reaction.}
\label{fig3}
\end{minipage}
\hspace{\fill}
\begin{minipage}[t]{0.46\textwidth}
\includegraphics[width=0.9\textwidth]{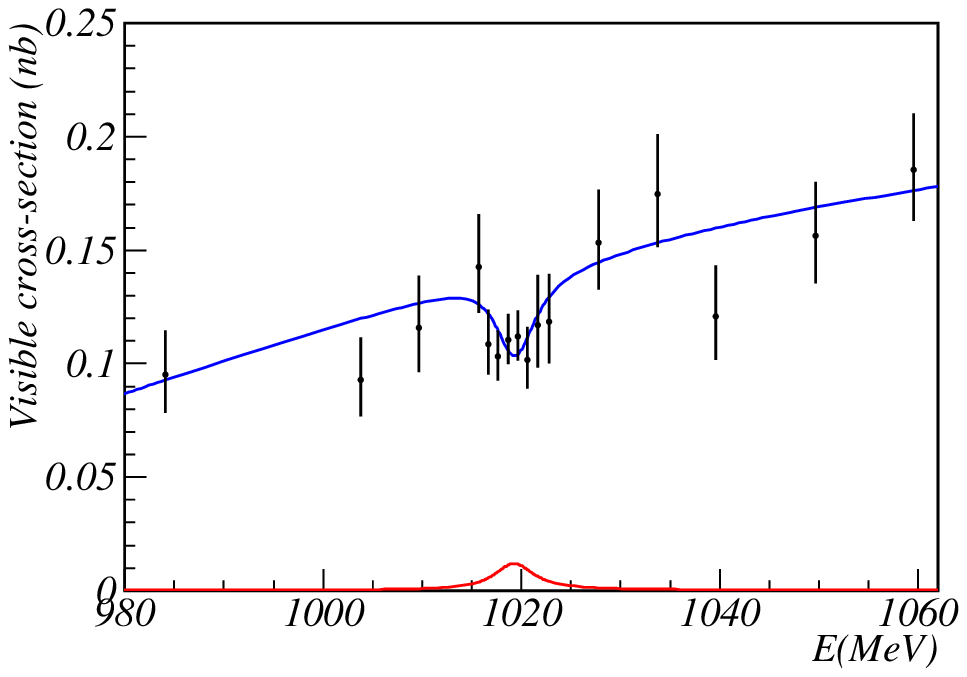}
\caption{The visible cross section of
$e^+e^-\to\omega\pi^0\to\pi^0\pi^0\gamma$ reaction.}
\label{fig4}
\end{minipage}
\end{figure}

 The decay $\phi\to\omega\pi$ was studied in two main $\omega$ decay
modes: $\pi^+\pi^-\pi^0$ and $\pi^0\gamma$.
The results of analysis of PHI96 data were first measurement of the probability
of this decay  in $4\pi$ final state\cite{ompic} and observation of
the interference pattern in $2\pi^0\gamma$ final state\cite{ompin}.
In this report we present
results of PHI98 experiment. The visible cross section for
$e^+e^-\to\omega\pi\to\pi^+\pi^-\pi^0\pi^0$ process is shown in
Fig.\ref{fig3}. The measured real and imaginary parts of
interference amplitude and corresponding branching ratio are following:
$$Re(Z)=0.112\pm0.015,\;Im(Z)=-0.104\pm0.022,$$
$$B(\phi\to\omega\pi^0)=(4.6\pm1.0\pm0.6)\cdot 10^{-5}.$$
This result can be compared with our first
measurement:$B(\phi\to\omega\pi^0)=(4.8^{+1.9}_{-1.7}\pm0.8)\cdot 10^{-5}$
\cite{ompic}. The value of interference amplitude is sensitive to mechanism
of $\phi-\rho$ and $\phi-\omega$ mixing. The VDM prediction with conventional
mixing of vector mesons
$B(\phi\to\omega\pi^0)=(8\div9)\cdot 10^{-5}$\cite{ompith}, that is about two
times more than measured value. 
Possible explanations of this deviation could be the nonstandard mixing
or the existence of direct $\phi\to\omega\pi^0$ transition.

The interference pattern for
$e^+e^-\to\omega\pi\to\pi^0\pi^0\gamma$ process (Fig.\ref{fig4}) has
another shape:
$$Re(Z)=0.02\pm0.03,\;Im(Z)=-0.24\pm0.05$$
In this final state there is additional contribution
of $\phi\to\rho^0\pi^0\to\pi^0\pi^0\gamma$ transition
which interfere with $e^+e^-\to\omega\pi^0$ amplitude. The difference
between interference amplitudes measured in $2\pi^0\gamma$ and $4\pi$ 
modes is in a qualitative agreement with VDM estimation for 
$\phi\to\rho^0\pi^0$ transition\cite{ompin}: $Re(Z)=-0.079,\;Im(Z)=-0.053$.

Detail description of study of $\phi\to\pi^+\pi^-$ decay is in report presented
at this conference by S.Burdin. The result is
$Re(Z)=0.061\pm0.005$, $Im(Z)=-0.042\pm0.06$,
$B(\phi\to\pi^+\pi^-)=(0.71\pm0.10\pm0.10)\cdot 10^{-4}$.
\section{$\rho$ and $\omega$ decay}
 The $e^+e^-\to\eta\gamma$ reaction was studied in energy region of $\rho$
and $\omega$ resonances in 7 photons final state. This state was chosen due to
absence of any physical background. The preliminary results of this study 
$$B(\omega\to\eta\gamma)=(5.9\pm1.0)\cdot 10^{-4},\;
B(\rho\to\eta\gamma)=(2.0\pm0.4)\cdot 10^{-4}.$$
have statistical accuracy comparable or better then world average one. 
The work is in progress.
\begin{figure}[tb]
\begin{minipage}[t]{0.46\textwidth}
\includegraphics[width=0.9\textwidth]{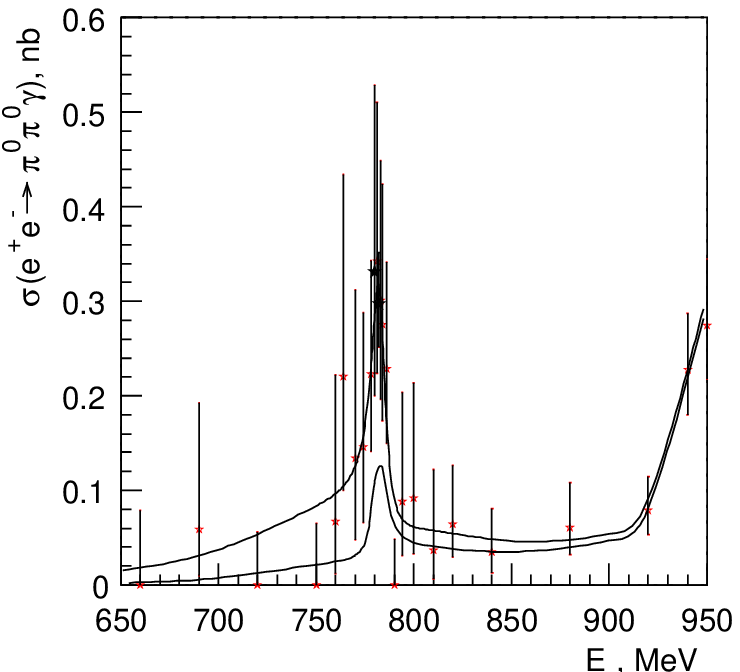}
\caption{The Born cross section of $e^+e^-\to\pi^0\pi^0\gamma$ process.
Points -- data. Lower curve -- VDM calculation. Upper curve -- fit result.}
\label{fig5}
\end{minipage}
\hspace{\fill}
\begin{minipage}[t]{0.46\textwidth}
\includegraphics[width=0.9\textwidth]{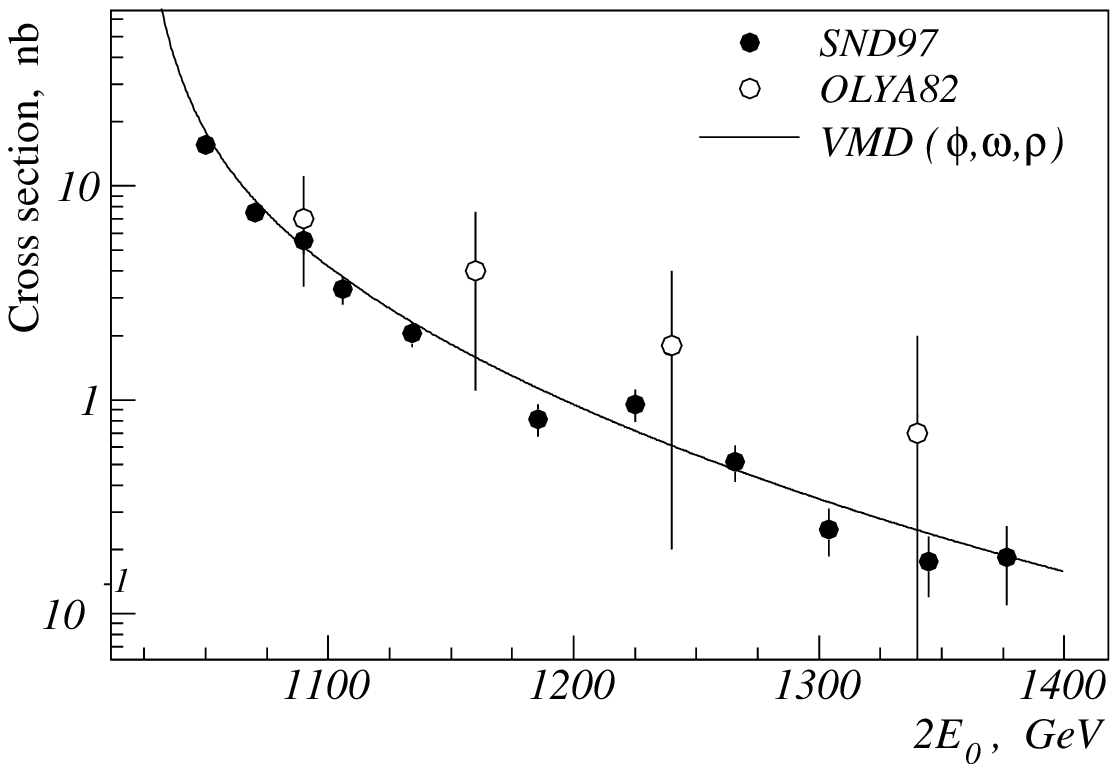}
\caption{The Born cross section of $e^+e^-\to\ K_S K_L$ process.
Points -- data. Curve -- the result of VDM calculation with $\rho$, $\omega$
and $\phi$ mesons. }
\label{fig6}
\end{minipage}
\end{figure}

The experimental data on $e^+e^-\to\pi^0\pi^0\gamma$ process in $\rho$,
$\omega$ mesons energy region are poor. Only one measurement of
$\omega\to\pi^0\pi^0\gamma$ decay was done:
$B(\omega\to\pi^0\pi^0\gamma)=(7.2\pm 2.5)\cdot 10^{-5}$\cite{ppgex}.
In VDM this process is described by sum of two amplitudes:
$\rho\to\omega\pi^0,\,\omega\to\pi^0\gamma$ and
$\omega\to\rho\pi,\,\rho\to\pi^0\gamma$. The VDM estimation 
$B(\omega\to\pi^0\pi^0\gamma)=2.6\cdot 10^{-5}$ contradicts to the
experimental value. There are no works explaining this 
discrepancy.

 About 300 events of $e^+e^-\to\pi^0\pi^0\gamma$ process were selected
in the OME98 experiment. The measured cross section together
with the result of VDM calculation are shown in Fig.\ref{fig5}. The
experimental data significantly exceeds the calculated curve. The measured
cross section was fitted by sum of amplitudes of $\rho\to\omega\pi^0$,
$\omega\to\pi^0\pi^0\gamma$ and $\rho\to\pi^0\pi^0\gamma$ decays.
We check $\rho^0\pi^0$, LIPS, $\sigma(700)\gamma$ models for description of
$\omega$ decay and LIPS, $\sigma(700)\gamma$ for $\rho$ decay. The analysis
of energy and angular distributions did not allow to choose any decay model
on our statistical level. The model dependence of the fit parameters
was included into systematic errors.
The following branching ratios was obtained as the fit result:
$$BR(\omega\to\pi^0\pi^0\gamma )=(8.4^{+4.9}_{-3.1}\pm3.5)\cdot 10^{-5},\;
BR(\rho\to\pi^0\pi^0\gamma )=(4.2^{+2.9}_{-2.0}\pm1.0)\cdot 10^{-5}.$$
The branching ratio of $\rho$ decay does not include $\omega\pi^0$ 
intermediate state contribution.  

The obtained probability of $\omega\to\pi^0\pi^0\gamma$ decay
significantly exceeds VDM predictions and is in a good agreement with
result of previous measurement. The relatively large value of
the $\rho\to\pi^0\pi^0\gamma$ branching ratio also can not be explained in
VDM framework. We consider this result as the first evidence of
$\rho\to\pi^0\pi^0\gamma$ decay.
\section{Hadron production cross sections}
The main goal of MHAD97 experiment was study of hadron
production cross sections. In energy region of MHAD97 experiment
the level of hadron cross sections is mainly defined by contribution of
the excited $\rho$ and $\omega$ state, parameters of which are not well
established.

The good reconstruction of $K_S$ meson from four photons allowed to select
events of $e^+e^-\to K_S K_L,\, K_S\to 2\pi^0$ process. The measured cross
section of this process is shown in Fig.\ref{fig6}. In the energy region
below 1.4 GeV the contributions of only $\rho$, $\omega$ and $\phi$ mesons
describe the cross section well.

\begin{figure}[tb]
\begin{minipage}[t]{0.46\textwidth}
\includegraphics[width=0.9\textwidth]{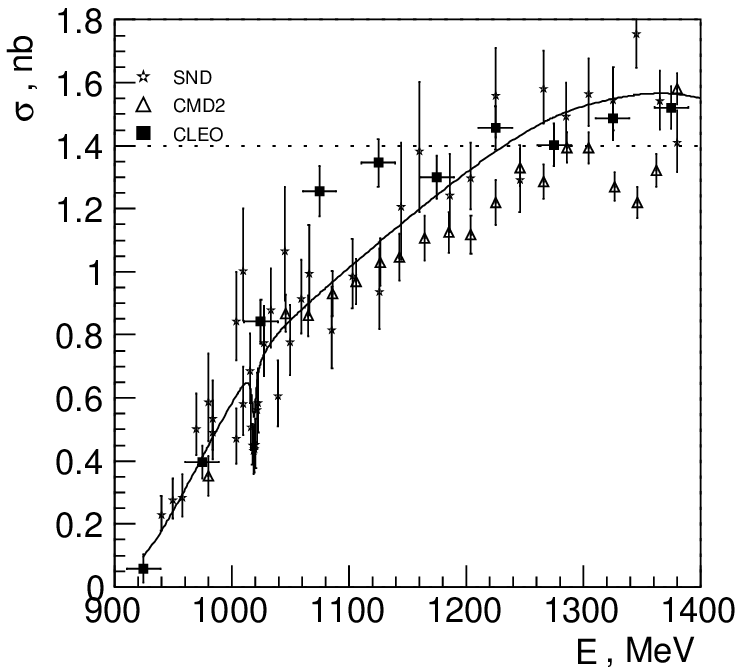}
\caption{The Born cross section of $e^+e^-\to\omega\pi^0\to\pi^0\pi^0\gamma$
process.}
\label{fig7}
\end{minipage}
\hspace{\fill}
\begin{minipage}[t]{0.46\textwidth}
\includegraphics[width=0.9\textwidth]{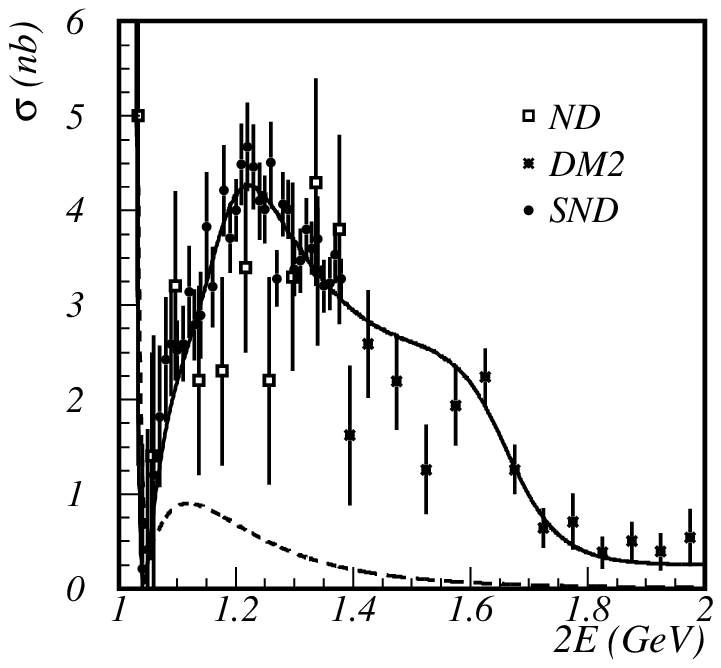}
\caption{The Born cross section of $e^+e^-\to\pi^+\pi^-\pi^0$ process.}
\label{fig8}
\end{minipage}
\end{figure}

The $e^+e^-\to\omega\pi^0$ cross section was measured in neutral mode from
threshold to maximum VEPP2-M energy. Obtained cross section together with 
results of other experiments is presented in Fig.\ref{fig7}. One can see
that there are systematic difference between results of different experiments.
Our systematic error is not more then 10\%.

The measured cross section of $e^+e^-\to\pi^+\pi^-\pi^0$\cite{mh3p} is shown in
Fig.\ref{fig8}. The broad peak in cross section is seen near 1.2 GeV and
there is no visible confirmation of the existence of the 
$\omega(1420)$\cite{PDG} state. We fit SND, CMD2 and DM2 data by sum of 
contributions of $\phi$,
$\omega$ and two excited states: $\omega^\prime$ and $\omega^{\prime\prime}$.
The Breit-Wigner shape
with constant width was used for description of excited $\omega$.
The parameters $\omega^{\prime\prime}(1600)$ were obtained from
DM2 data on the reaction $e^+e^-\to\omega\pi^+\pi^-$.
The best fit of experimental data was obtained with the following
resonance phases choice: $\phi_{\omega}=0$, $\phi_{\omega^\prime}=\pi$ and
$\phi_{\omega_{\prime\prime}}=0$. The resulting curve is shown
in Fig.\ref{fig8}. The mass about 1200 MeV was obtained for $\omega^\prime$
state instead of table 1420 MeV value. The inclusion of mass-dependent
widths of $\omega^\prime$ and $\omega^{\prime\prime}$ into the fit can
significantly shift
$\omega^\prime$ position from visible one. But in any case the
reexamination of table $\omega(1420)$\cite{PDG} state based on analysis of 
all available data are required.
\section{Acknowledgment}
The authors are grateful to N.N.Achasov for useful discussions.
The work is partially supported by RFBR (Grants No 96-15-96327, 99-02-17155,
99-02-16815, 99-02-16813) and STP ``integration'' (Grant No 274).

\end{document}